\documentclass[aps,amsmath,amssymb]{revtex4}
\usepackage{psfig}
\usepackage{graphicx}

\begin{document}

\title{\bf The difficulty in the decay constants and spectra of $D^{*}_{sJ}$}
\date{\today}
\author{\bf  Ron-Chou Hsieh$^{a}$\footnote{Email:
hsieh@hepth.phys.ncku.edu.tw},Chuan-Hung
Chen$^{a}$\footnote{Email: physchen@mail.ncku.edu.tw} and
Chao-Qiang Geng$^{b}$\footnote{Email: geng@phys.nthu.edu.tw} }

\vskip1.0cm

\affiliation{$^{a}$Department of Physics, National Cheng-Kung
University, Tainan 701, Taiwan \\
$^{b}$Department of Physics, National Tsing-Hua University,
Hsin-Chu , Taiwan}

\begin{abstract}
We examine the compatibility in predicted masses and the decay
constants of $D^{*}_{sJ}$ mesons in terms of two-quark contents.
We find that the results of a specific model, which is governed by
heavy quark limit, will encounter a challenge to fit both spectra
and decay constants simultaneously.
\end{abstract}


\maketitle

In spite of the success of the quark model, the explanation
with the same concept on  light scalar mesons denoted by $0^{++}$,
such as the nonet composed of isoscalars $\sigma(600)$ and
$f_{0}(980)$, isovector $a_{0}(980)$ and isodoublet $\kappa$,
still has  puzzles of (a) why $a_{0}(980)$ and $f_0(980)$ are
degenerate in masses and (b) why the widths of $\sigma$ and
$\kappa$ are broader than those of $a_{0}(980)$ and $f_{0}(980)$
\cite{Cheng}. Probably, these scalar states consist of four
rather than two quarks \cite{4q}. Moreover, the possibilities of
$K\bar{K}$ molecular states and scalar glueballs are also
proposed. Hence, the conclusion is still uncertain.

The mysterious event happens not only in the light scalar mesons,
but now also in the heavy charm-quark system. Recently, BABAR
collaboration has observed one narrow state, denoted by
$D^{*}_{sJ}(2317)$, from a $D^{+}_{s} \pi^{0}$ mass distribution
\cite{Babar1}. Afterward, the same state is confirmed by CLEO and
a new state $D^{*}_{sJ}(2460)$ is also seen in the $D^{*+}_{s}
\pi^0$ final state \cite{CLEO}. Finally, BELLE verifies the
observations \cite{Belle1}. The two states $D^{*}_{sJ}(2317)$ and
$D^{*}_{sJ}(2460)$ currently are identified as parity-even states
with ${\bf 0}^{+}$ and ${\bf 1}^{+}$, respectively. According to
the experimental displays, the masses (widths) of both states are
too low (narrow) and cannot match with theoretical predictions
\cite{Theory}. To explain the discrepancy, either the theoretical
models have to be modified \cite{BEH} or the observed states are
the new composed states. To satisfy the latter, many interesting
solutions have been suggested recently in Refs.
\cite{CJ,BCL,BR,YH,Szczepaniak,CD,Chen-Li}.

It is interesting to speculate whether a two-quark picture is
really enough to explain the spectra of the $D^{*}_{sJ}$ system.
Before examining the possibility, we first discuss the nonstrange
parity-even states $D^{**}$, measured by CLEO
\cite{CLEO-96} and BELLE \cite{Belle-conf} in $B$ meson decays
before the BABAR's observation. With the two-quark picture, there
are four parity-even (angular momentum $\ell=1$) states described
by $J^P=0^{+}$, $1^{+}$, $1^{+}$ and $2^{+}$, respectively.
Here, $J=j_{q}+S_{Q}$ is the total angular momentum of the corresponding
meson and consists of the angular momentum of the light quark,
$j_{q}$, and the spin of the heavy quark, $S_{Q}$, where
$j_{q}=S_{q}+\ell$ is combined by the spin and orbital angular
momenta of the light quark. In the literature, they are usually
labeled by $D^{*}_{0}$, $D^{\prime}_{1}$, $D_{1}$ and
$D^{*}_{2}$, respectively.  The first two belong to $j_{q}=1/2$,
while the last two are $j_{q}=3/2$. In the heavy quark limit, it
is known that $D^{*(\prime)}_{0(1)}$ and $D^{(*)}_{1(2)}$ belong
to the doublets ($0^{+}$, $1^+$) and ($1^{+}$, $2^+$),
respectively \cite{Cheng-hep}. Moreover, the former decay only via
$S$-wave while the latter are through $D$-wave. Therefore, one
expects that the widths of the former are much broader than those
of the latter, which is consistent with the observations of CLEO
and BELLE \cite{CLEO-96,Belle-conf}. Even the BELLE's updated data
\cite{Belle-new} also show the same phenomenon. We now summarize
the results of CLEO and BELLE as follows. In CLEO, the masses
(widths) of $P$-wave states are given by
$m_{D_{1}}(\Gamma_{D_{1}})=2422.0 \pm 2.1\; (18.9^{+4.6}_{-3.5})$
and $m_{D^{*}_{2}}(\Gamma_{D^{*}_{2}})=2458.9 \pm 2.0\; (23\pm 5)$
MeV. In BELLE, the four states are all measured as
$m_{D^*_{0}}(\Gamma_{D^{*}_{0}})=2308 \pm 17 \pm 15 \pm 28\; (276
\pm 21 \pm 18 \pm 60)$,
$m_{D^{\prime}_{1}}(\Gamma_{D^{\prime}_{1}})=2427.0 \pm 26 \pm 20
\pm 15\; (384^{+107}_{-75}\pm 24 \pm 70)$,
$m_{D_{1}}(\Gamma_{D_{1}})=2421.4 \pm 1.5 \pm 0.4 \pm 0.8\; (23.7
\pm 2.7\pm 0.2 \pm 4.0)$, and
$m_{D^{*}_{2}}(\Gamma_{D^{*}_{2}})=2461.6 \pm 2.1 \pm 0.5 \pm
3.3\; (45.6 \pm 4.4 \pm 6.5 \pm 1.6)$ MeV. Clearly, from BELLE's
results, the predicted mass differences $M(D^{*}_{0})-M(D)\, [
M(D^{\prime*}_{1})-M(D^*)]\sim 350$ MeV in the limit of the heavy
quark symmetry \cite{BEH}, in which $D(D^*)$ belong to the doublet
($0^-$, $1^-$), are
smaller than
experimental results $\sim 420$ MeV. As mentioned in Ref.
\cite{BEH}, the correction $\Lambda_{QCD}/m_{c}$ will compensate
for the shortage. Hence, if the parity-even nonstrange charm-meson
obeys the chiral quark model, it is incomprehensible
intuitively why the strange one cannot do.

In fact, there is somewhat a difference in the decays of
$D^{*}_{sJ}$ and $D^{**}$. Since the measured masses of
$D^{*}_{sJ}$ are just below the $D^{(*)}K$ threshold and the
corresponding widths are at the limit of the detector, being less than
$10$ MeV, both parity-even mesons could only decay through isospin
violating channels to $D\pi$ and $D^* \pi$. By this reason, one
would understand why the widths of the parity-even mesons are so
narrow. As a result, in terms of the pole of the Breit-Wigner formula,
the effective masses of the propagating mesons while they are
produced are estimated by $M_{eff}\sim (
M^{2}(D^{*}_{sJ})+M(D^{*}_{sJ})\Gamma(D^{*}_{sJ}))^{1/2}\approx
M(D^{*}_{sJ})$. That is, there is no room in phase space for
$D^{*}_{sJ}$ decaying to the final states $D^{(*)}K$. The states
in the $\bar{c} s$ system should be highly deformed inside the
charm-meson compared to conventional $0^{-}(1^{-})$ states. On the
contrary, there is no any suppression rule on  $D^{*}_{0}
\to D \pi$ or $D^{\prime}_{1}\to D^* \pi$ so that the masses
(widths) are not far away from our expectations. Therefore, it is
important to find out a proper method to explain the observed
states $D^{*}_{sJ}$ in the framework of the quark model.

Recently,  Deandrea {\it et. al.} \cite{CQM1} have showed
that in terms of the constituent quark meson (CQM) model
\cite{CQM2,CQM3}, the masses of $D^{*}_{sJ}$ could be constructed
to be consistent with the measurements at $B$ factories. It is
interesting to question whether the CQM model can really address
the other suffering problems in $D^{*}_{sJ}$ mesons. In order to
uncover the doubt, we calculate the decay constant of
$D^{*}_{s0}(2317)$, defined by
\begin{eqnarray}
\langle 0| \bar{c} \gamma^{\mu} s| D^{*}_{s0}(2317)>=i p^{\mu}
f_{D^{*}_{s0}},\label{fd}
\end{eqnarray}
in the CQM model and make a comparison to
$f_{D_{s}}$. Under the heavy quark limit, we also assume
$f_{D^*_{s1}}(2460)\sim f_{D^*_{s0}}(2317)$. Since the CQM model is
based on the bosonized Nambu-Jona-Lasinio model and heavy quark
symmetry, the constituent quark mass should be
governed by the Schwinger-Dyson equation 
\begin{eqnarray}
m&=&m_{0}+8mG_{1} I_{1}, \\
I_{1}&=&\frac{iN_{c}}{16\pi^4} \int^{reg} d^{4}k
\frac{1}{k^2-m^2}=\frac{N_{c}m^{2}}{16\pi^2}
\Gamma\left(-1,\frac{m^{2}}{\Lambda^2},\frac{m^{2}}{\mu^2}\right),\nonumber
\end{eqnarray}
where $G_{1}$ is taken to be as a parameter and flavor
independent, $m_{0}$ is the current quark mass, $N_{c}=3$ is the
color number, $\Lambda$ is the ultraviolet (UV) cutoff, which is
the scale to separate the degrees of quark freedom  from those at
hadron level, and $\mu$ is the infrared (IR) cutoff, which is used
to interrupt the confining effects. Furthermore, the chosen values
of $\Lambda$ and $\mu$ have to satisfy the condition
$\Delta=M-m_{Q}\geq m$ to guarantee no effects from confinement,
in which $M$ and $m_{Q}$ correspond to the heavy meson and quark.
At the heavy quark symmetry limit, it is known that the doublet
($0^-, 1^{-}$) can be related to ($0^+, 1^{+}$), {\it i.e.}, one
can obtain $\Delta_{S}$ from $\Delta_{H}$ where $H(S)$ denote
parity-odd (even) multiplet. Hence, according to the result of
Ref. \cite{CQM3}, the relationship for $\Delta_{H}$ and
$\Delta_{S}$ is given by $\Pi(\Delta_{H})=\Pi(\Delta_{S})$ with
\begin{eqnarray}
\Pi(\Delta_{H,S})&=&I_{1}+(\Delta_{H,S}\pm m)I_{3}(\Delta_{S}),
\\ I_{3}(\Delta)&=&\frac{N_{c}}{16\pi^{3/2}}
\int^{1/\mu^{2}}_{1/\Lambda^2} ds
\frac{e^{-s(m^2-\Lambda^2)}}{s^{3/2}}(1+
erf(\Lambda\sqrt{s})),\nonumber
\end{eqnarray}
where the $\Pi$ function comes from the meson self energy.
Consequently, the renormalization constant of the meson could be
obtained via $Z^{-1}(x)=(d\Pi(x)/dx)_{x=\Delta}$ and described by
\begin{eqnarray}
Z^{-1}_{H,S}=(\Delta_{H,S}\pm m)\frac{\partial
I_{3}(\Delta_{H,S})}{\partial \Delta_{H,S}}+I_{3}(\Delta_{H,S}).
\end{eqnarray}

\begin{figure}[h]
\includegraphics*[width=2.5
in]{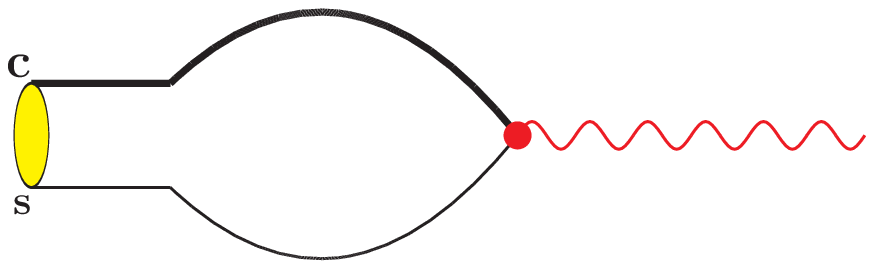}
\caption{Diagram for weak decay of $D^{*}_{sJ}$} \label{fey}
\end{figure}

After introducing the basics formulas
in the CQM model,
the decay constant $f_{D^*_{sJ}}(2317)$ can be calculated in terms
of the definition in Eq. (\ref{fd}). To be more clear, the
corresponding diagram is shown in Fig. \ref{fey}.
Hence, the explicit expression is written as \cite{CQM3}
\begin{eqnarray}
f_{D^*_{s0}}(2317)=\frac{2}{\sqrt{m_{D^*_{s0}}}}
\sqrt{Z_{S}}\Pi(\Delta_{S}).
\end{eqnarray}
To obtain the numerical value, we adopt $G_{1}=5.25$ GeV$^{-2}$,
$\mu=m=0.5$ and $\Lambda=1.25$ GeV which  are used in Ref.
\cite{CQM1} to reach the right spectra of $D^*_{sJ}$. Immediately,
we get $f_{D^{*}_{s0}}\approx 176$ MeV. With the similar process,
the decay constant of $D_{s}$ is also estimated to be $f_{D_{s}}
\approx 270$ MeV. Is the obtained value of $f_{D^{*}_{sJ}}$ in the CQM
model proper although the value of $f_{D_{s}}$ is still consistent with
that in particle data group (PDG) \cite{PDG}? We could examine
this validity by the BELLE's observations in $B$ decays. BELLE
recently reports the decay branching ratios of  $B\to
D^{*}_{sJ} \bar{D}$ to be \cite{Belle2}
\begin{eqnarray*}
BR(B^{+} \to D^{*+}_{sJ}(2317) \bar{D}^0) \times BR(D^{*+}_{sJ}
\to D^{+}_{s} \pi^{0})&=&(8.1^{+3.0}_{-2.7}\pm 2.4)\times 10^{-4}, \\
BR(B^{+} \to D^{*+}_{sJ}(2317) \bar{D}^0) \times BR(D^{*+}_{sJ}
\to D^{*+}_{s} \gamma)&=&(2.5^{+2.1}_{-1.6})(<7.6)\times 10^{-4}, \\
BR(B^{+} \to D^{*+}_{sJ}(2460) \bar{D}^0) \times BR(D^{*+}_{sJ}
\to D^{+*}_{s} \pi^{0})&=&(11.9^{+6.1}_{-4.9}\pm 3.6)\times 10^{-4}, \\
BR(B^{+} \to D^{*+}_{sJ}(2460) \bar{D}^0) \times BR(D^{*+}_{sJ}
\to D^{+}_{s} \gamma)&=&(5.6^{+1.6}_{-1.5} \pm 1.7)\times
10^{-4}, \\
BR(B^{+} \to D^{*+}_{sJ}(2460) \bar{D}^0) \times BR(D^{*+}_{sJ}
\to D^{+*}_{s} \gamma)&=&(3.1^{+2.7}_{-2.3} )(<9.8)\times 10^{-4}.
\end{eqnarray*}
By assuming that $D^{*}_{sJ}$ mainly decay to $D^{(*)} \pi$ and
$D^{(*)} \gamma$ and taking the central values of measured
results, the BRs of the decays $B^{+}\to D^{*+}_{s0}(2317)
\bar{D}^{0}$ and $B^{+}\to D^{*+}_{s1}(2460) \bar{D}^{0}$ are
roughly estimated to be $10.6 \times 10^{-4}$ and $20.6 \times
10^{-4}$, respectively. Since the decays $B^{+}\to D^{*+}_{sJ}
\bar{D}^{0}$ are
color-allowed processes, the nonfactorizable effects are
negligible. Therefore, comparing to the BRs of $B^{+}\to D^{+}_{s}
\bar{D}^{0}$ and $B^{+}\to D^{*+}_{s} \bar{D}^{0}$ in PDG
\cite{PDG}, we have
\begin{eqnarray}
R_{1}&\equiv&{BR(B^{+}\to D^{*+}_{s0}(2317) \bar{D}^{0}) \over
BR(B^{+}\to D^{+}_{s}
\bar{D}^{0})}=\frac{f^{2}_{D^{*}_{s0}}}{f^{2}_{D_{s}}}\sim 0.10, \label{r1}\\
R_{2}&\equiv&{BR(B^{+}\to D^{*+}_{s1}(2460) \bar{D}^{0}) \over
BR(B^{+}\to D^{*+}_{s}
\bar{D}^{0})}=\frac{f^{2}_{D^{*}_{s1}}}{f^{2}_{D^{*}_{s}}}\sim
0.23. \label{r2}
\end{eqnarray}
With the previous obtained value of $f_{D^{*}_{s0}}\approx 176$
MeV, we clearly see that $R_{1CQM}\sim R_{2CQM} \sim 0.42$ are much
bigger than experimental requirements. In order to reduce the values
of $f_{D^*_{sJ}}$, it seems that we should adopt different values
of the parameters $\Delta_{H,S}$, $\Lambda$ and $\mu$. For
simplicity, we still fix $\Lambda$ to be $1.25$ GeV. The predicted
results in the CQM model with various values of $m_{s}$ and $\mu$
by using $\Delta_{H}=m_{s}$ are displayed in Table \ref{fd1}.
\begin{table}[htb]
\caption{Decay constants $(f_{D^{*}_{s0}}, f_{D_{s}})$ with
various values of $m_{s}$ and $\mu$ by fixing
$\Delta_{H}=m_{s}$.}\label{fd1}
\begin{ruledtabular}
\begin{tabular}{cccc}
$\mu \setminus m_{s}$ (GeV) & $0.5 $ & $0.55$ & $0.6$  \\\hline
0.35 & $(0.154,\ 0.262)$ & $(0.154,\ 0.267)$ & $(0.154,\ 0.272)$\\
0.25 & $(0.123,\ 0.246)$ & $(0.122,\ 0.250)$ & $(0.122,\ 0.253)$\\
\hline
\end{tabular}
\end{ruledtabular}
\end{table}
 From the table, we find that if we take $\mu=0.35$ GeV, the
ratio defined in Eq. (\ref{r1}) will be $R_{1,2}\sim 0.33$.
However, if we use $\mu=\Lambda_{QCD}=0.25$ GeV, the ratio $R_{1,
2CQM}$ could be around $0.24$, which is still large for $R_{1}$
although it fits $R_{2}$ well. Nevertheless, according to the
analysis of Ref. \cite{CQM1}, the calculated
$\Delta_{S}-\Delta_{H}\sim 210$ MeV implies that the average mass
of the parity-even meson, defined as
$M_{S}=(3M_{D^{*}_{s1}}+M_{D^*_{s0}})/4$, is around $2286$ MeV. By
taking $M_{D^*_{s0}}=2317$ MeV, we get $M_{D^{*}_{s1}}=2275$ MeV
which is even smaller than $M_{D^*_{s0}}$. It seems that if we
just concentrate on the formulas of  the heavy quark limit in the CQM
model, the suffering problems, low mass and large decay constant,
cannot be solved.

In summary, we have studied the spectra and decay constants of
$D^{*}_{sJ}$ in the CQM model simultaneously. We have shown that
the chosen values of the parameters for fitting the corrected mass
of $D^{*}_{sJ}$ will give too big decay constants $f_{D^*_{sJ}}$,
which are constrained by the observed decays $B\to D^{*}_{sJ}
\bar{D}$. We note that the predicted decay constants on
$D^{*}_{sJ}(2317)$ and $D^{*}_{sJ}(2460)$ are $84$ and $126$ MeV
in the covariant light-front QCD \cite{CCH}, respectively. We
conclude that the conventional approach is difficult to fit both
spectra and decay constants of $D^{*}_{sJ}$.
\\

\noindent {\bf Acknowledgments}

This work is supported in part by the National Science Council of
R.O.C. under Grant No. NSC-91-2112-M-001-053 and No.
NSC-92-2112-M-006-026.

\end{document}